\documentclass[11pt,a4paper]{scrartcl}

\usepackage{ILD}
\usepackage[numbers,sort&compress]{natbib}
\usepackage[symbol]{footmisc}
\usepackage{feynmf}
\usepackage{multirow}
\usepackage{textpos}
\usepackage{lineno}
\usepackage{afterpage}

\usepackage{tabularx}
    \newcolumntype{L}{>{\raggedright\arraybackslash}X}
\def\epsilonb{\ensuremath{\epsilon_{b}}\xspace}
\def\epsilonc{\ensuremath{\epsilon_{c}}\xspace}
\def\epsilonuds{\ensuremath{\epsilon_{uds}}\xspace}
\def\epsilonb2{\ensuremath{\epsilon^{2}_{b}}\xspace}
\def\epsilonc2{\ensuremath{\epsilon^{2}_{c}}\xspace}
\def\epsilonuds2{\ensuremath{\epsilon^{2}_{uds}}\xspace}

\def\costheta{\ensuremath{\cos \theta}\xspace}

\def\ab{ab\ensuremath{^{-1}}\xspace}

\def\ssbar{\ensuremath{s}\ensuremath{\overline{s}}\xspace}
\def\qqbar{\ensuremath{q}\ensuremath{\overline{q}}\xspace}

\def\eLpR{\ensuremath{e_L^{-}e_R^{+}}\xspace}
\def\eRpL{\ensuremath{e_R^{-}e_L^{+}}\xspace}

\def\dEdx{\ensuremath{dE/\/dx}\xspace}
\def\dNdx{\ensuremath{dN/\/dx}\xspace}

\def\AFB{\ensuremath{A_{FB}}\xspace}

\def\Pb{\ensuremath{P_{chg.}}\xspace}

\title{$e^+e^- \rightarrow s\bar{s}$ at $\sqrt{s} = 250$ GeV at future linear colliders}
\titlecomment{This work was carried out in the framework of the ILD concept group}
\titlecomment{Talk presented at the International Workshop on Future Linear Colliders (LCWS 2025)}

\ildproc{PHYS}{2026}{008} % Proceedings% Example:
\date{\today}
\addauthor{J.P. M\'arquez}{\institute{1} 
\footnote{Speaker \href{mailto:jesus.marquez-hernandez@ijclab.in2p3.fr}{jesus.marquez-hernandez@ijclab.in2p3.fr}}}
\addauthor{R. P\"oschl}{\institute{1}} 
\addauthor{A. Irles}{\institute{2}}
\addauthor{F. Richard}{\institute{1}}
\addinstitute{1}{IJCLab, CNRS \& U. Paris-Saclay, 15 rue Georges Clémenceau, Orsay, 91400, France}
\addinstitute{2}{IFIC, Universitat de València \& CSIC, 2 Catedrático José Beltrán, Paterna, 46980, Spain}

\abstract{
The forward--backward asymmetry (\AFB) in light-quark production is a sensitive probe of the electroweak sector and potential flavor-dependent BSM effects. We present a study of $e^+e^- \rightarrow s\bar{s}$ at $\sqrt{s}=250$~GeV at future linear colliders, using full International Large Detector (ILD) simulation and reconstruction tools for the International Linear Collider (ILC), and the Linear Facility at CERN (LCF@CERN). 

We assess the impact of particle identification on charge reconstruction and \AFB extraction, considering software improvements using Comprehensive PID (CPID) for optimal \dEdx usage, as well as hardware scenarios including cluster counting (\dNdx) and ideal Time Projection Chamber (TPC) performance. Improvements in statistical discrimination are evaluated with corrections for charge misidentification and acceptance applied.

Results indicate that precise $A_{FB}^{s\bar{s}}$ measurements are feasible and that advanced PID is key to maximizing sensitivity to electroweak and new-physics effects.
}

\graphicspath{ {./logos/}{./figures/} }

\begin{document}

\titlepage

%\tableofcontents

\section{Introduction}
\label{sec:intro}

The study of quark pair production in $e^+e^-$ collisions provides a precise probe of electroweak interactions and is sensitive to the chiral structure of the SM, which could be modified by BSM contributions. This in  turn would lead to measurable modulations of the differential cross section that impact the \textit{Forward-Backward Asymmetry} (\AFB), defined as:

\begin{equation}
A^{q\bar{q}}_{FB}=\frac{\sigma^F - \sigma^B}{\sigma^F + \sigma^B}.
\end{equation}

Here, $\sigma^{F/B}$ is the differential cross-section integrated over the forward/backward hemisphere\footnote{Forward and backward stand for $\cos\theta > 0$, and $\cos\theta < 0$, respectively.}. This work focuses on $e^+e^- \to s\bar{s}$ at $\sqrt{s} = 250$ GeV, comparing ILC250~\cite{Behnke:2013xla,Baer:2013cma,Adolphsen:2013jya,Adolphsen:2013kya,Behnke:2013lya} and LCF@CERN250~\cite{LinearCollider:2025lya,LinearColliderVision:2025hlt}. This analysis assumes integrated luminosities of 2 \ab (ILC) and 3 \ab (LCF), applied to rescale statistical uncertainties. Four beam polarisation schemes are envisioned: P$(e^-,e^+)$=$(\pm0.8,\pm0.3)$. For simplicity, we denote P$(e^-,e^+)$=$(-0.8,+0.3)$ as \eLpR\ and $(+0.8,-0.3)$ as \eRpL. In this work, we consider only the \eLpR and \eRpL polarizations, each with 45\% of the total integrated luminosity.

Following previous heavy-quark studies~\cite{Irles:2019mdp,Okugawa:2019ycm,Irles:2020gjh,Irles:2023nee,Irles:2023ojs,Marquez:2023guo,Irles:2024ipg}, we extend the analysis to strange-quark production~\cite{Okugawa:2022zmt,Okugawa:2024dks}.

\section{Methodology and results in the baseline \dEdx analysis}
\label{sec:previousanalysis}
\subsection*{Preselection of \qqbar events}
The events are simulated (Refs.~\cite{Kilian:2007gr,Frank:2014zya,Schulte:1999tx,Sj_strand_2006}) and reconstructed with the standard ILD chain (Refs.\cite{Thomson_2009,Suehara:2015ura,ILD:2020qve}), producing \textit{Particle Flow Objects} (PFOs), jets and flavour-tagging information. This reconstruction uses LCFI+ (Ref.~\cite{Suehara:2015ura}) for jet clustering, vertexing, and flavor tagging. Jet clustering is done by using the VLC-algorithm (Ref.~\cite{Boronat:2016tgd}).

After reconstruction of the main physics objects (tracks, jets, PFOs, heavy quark tagging, and PID), an event preselection is applied using kinematic cuts to enrich the signal and reduce backgrounds. The backgrounds suppressed by this preselection arise from heavy boson pair production ($W^{+}W^{-}$, $ZH$, $ZZ$). The same set of cuts also mitigates the impact of radiative return events, which would otherwise dilute the measured cross section. The preselection at 250 GeV is as in Ref.~\cite{Irles:2023ojs}:
\begin{enumerate}
    \item Photons are vetoed by rejecting events where at least one of the following conditions is fulfilled:
    \begin{itemize}
        \item at least one jet contains only one PFO;
        \item at least one jet contains a reconstructed $\gamma_{\text{cluster}}$ with $E > 115$ GeV or located in the forward region $|\cos\theta|>0.97$;
    \end{itemize}
    \item Events with $\sin \Psi_{\text{acol}} > 0.3$ are rejected;
    \item Events with $m_{jj} < 140$ GeV are rejected;
    \item Events with $y_{23} > 0.02$ are rejected.
\end{enumerate}

In this preselection: $\gamma_{\text{cluster}}$ is an object that congregates all the neutral PFOs in a jet, $\sin \Psi_{\text{acol}}$ is the $\sin$ of the acollinearity of the jets ($\sin{\Psi_{acol}}=|\vec{p_{q}} \times \vec{p}_{\bar{q}}|/(|\vec{p_{q}}|\cdot|\vec{p}_{\bar{q}}|)$), $m_{jj}$ is the invariant mass of the two-jets system, and $y_{23}$ is the distance at which the clustering algorithm (VLC algorithm) changes from 2-jet into a 3-jet system.

\subsection*{Strange tagging based on kaon PID}
Until recently, no strange tagging algorithm was available. Therefore, a simple cut-based $s$-tagger was developed in Ref.~\cite{Okugawa:2024dks}. The cuts are defined using the existing b and c tagging, jet vertices, leading and subleading PFOS (LPFO, SPFO), track PID, and jet charge. The cut-flow of the $s$-tagging in this analysis is slightly different from that of the reference and is listed in Tab.~\ref{tab:cuts}.

\begin{table}[!ht]
\centering
\begin{tabular}{ccc}
\hline
\# & Name               & Quantity                               \\ \hline
1  & b-tag              & b-tag $<$ 0.3                          \\
2  & c-tag              & c-tag $<$ 0.65                         \\
3  & nvtx               & nvtx $= 1$                             \\
4  & LPFO momentum      & $p_{\text{LPFO}} > 15$ GeV              \\
5  & LPFO acollinearity & $\cos \theta_{\text{LPFO}_{1,2}} > 0.95$ \\
6  & Offset             & $V_0 = \sqrt{d_0^2 + z_0^2} < 1$ mm      \\
7  & \dEdx PID K        & If $|\cos \theta| < 0.85$, $|k_{\mathrm{dist}}| < 1.5$,  \\
   &                    & else $|k_{\mathrm{dist}}| < 0.5$        \\
8  & SPFO               & $p_{\text{SPFO}} < 10$ GeV and $Q_{\text{SPFO}} \neq Q_{\text{LPFO}}$ \\
9  & Charge             & $Q_{\text{LPFO}_1} \times Q_{\text{LPFO}_2} < 0$        \\
\hline
\end{tabular}
\caption{Summary of the event selection cuts used in the baseline \dEdx analysis.}
\label{tab:cuts}
\end{table}

The event selection cuts are grouped as follows: cuts~1--3 suppress heavy-quark backgrounds, 4--7 select s-jets via kaon PID. Cut 8 mitigates migration effects associated with sign flips in the \qqbar system direction by rejecting events that contain a high-momentum secondary PFO with charge opposite to that of the leading PFO, which could otherwise compete as the leading candidate. Cut 9 removes events where both jets have the same charge, inconsistent with a \qqbar final state and suggestive of misreconstruction.

\begin{figure}[!ht]
\centering
\begin{tabular}{ccc}
      \includegraphics[width=0.3\textwidth]{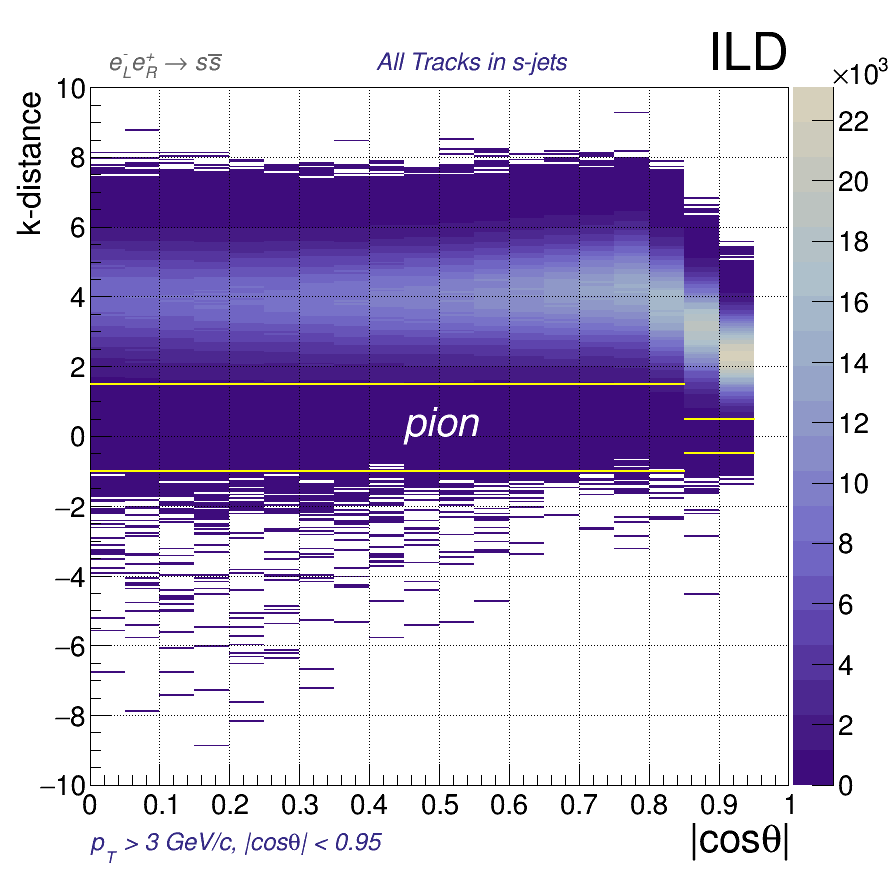} &
      \includegraphics[width=0.3\textwidth]{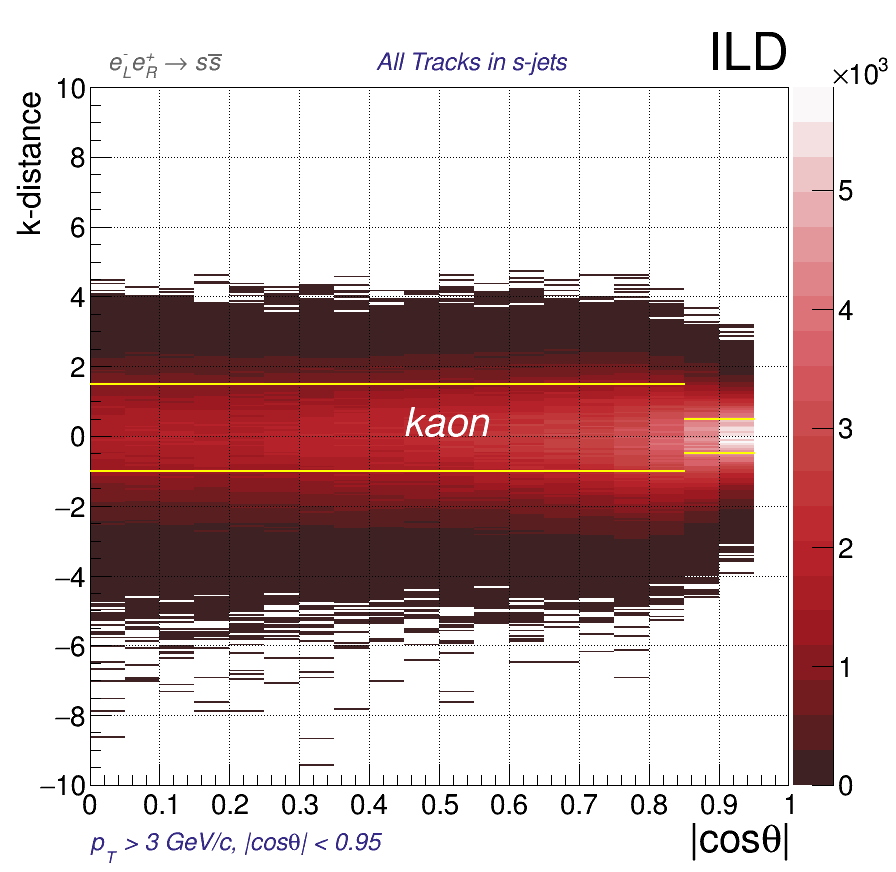} &
      \includegraphics[width=0.3\textwidth]{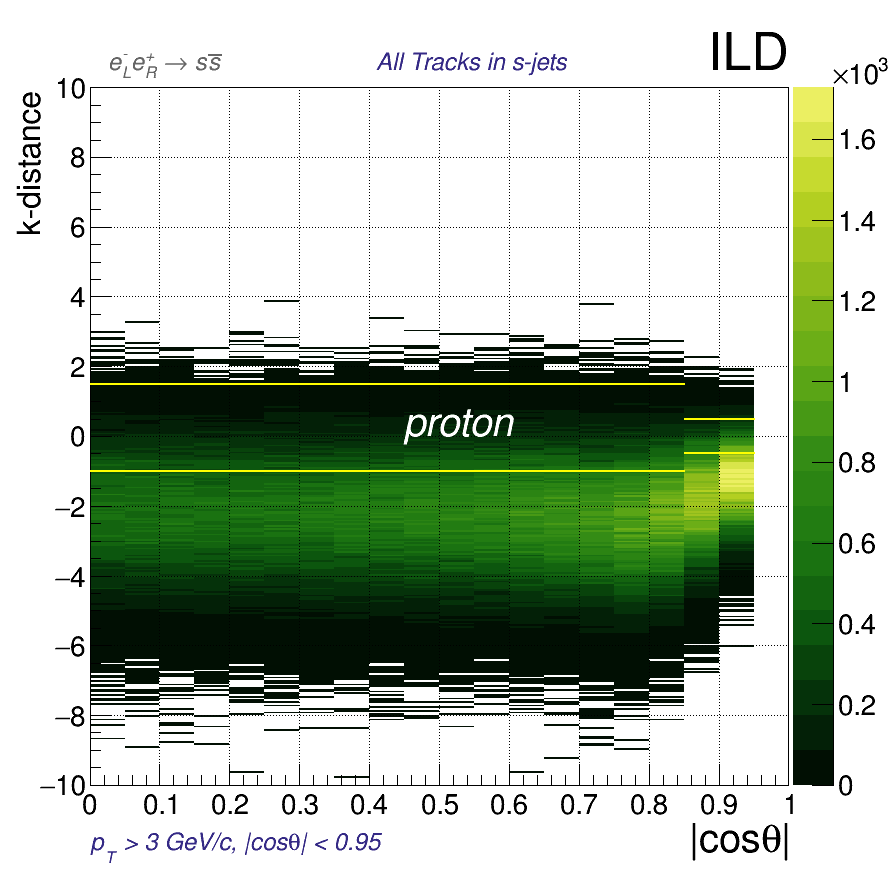}
    \end{tabular}
\caption{Kaon selection region in the $k$-distance plane used in cut~7.}
\label{fig:kdistcut}
\end{figure}

Compared to the analysis in Ref.~\cite{Okugawa:2024dks}, the main modification concerns cut~7. 
A two-dimensional mapping of the PID observables is used to define a region identifying a PFO as a kaon. 
The track \dEdx significance quantifies the compatibility of a track with a given particle hypothesis, and it is defined as $S_{\mathrm{d}E/\mathrm{d}x} = \Delta(\mathrm{d}E/\mathrm{d}x)/\sigma_{\mathrm{d}E/\mathrm{d}x}$, where $\Delta(\mathrm{d}E/\mathrm{d}x)=(\mathrm{d}E/\mathrm{d}x)_{\mathrm meas} - (\mathrm{d}E/\mathrm{d}x)_{\mathrm exp}$.
In this work, we take the Bethe-Bloch expectation for kaons as reference, which defines the \(\mathrm{d}E/\mathrm{d}x\) \textit{$k$-distance}, shown in Fig.~\ref{fig:kdistcut}.

For each of the nine cuts, the total efficiency and the efficiency as a function of \costheta are reported in the Appendix.

For the \AFB measurement, the jet charge is taken from the leading PFO, identified as a primary kaon from the $s$ or $\bar{s}$ quark. The selected sample is corrected sequentially: backgrounds are subtracted using templates scaled by their efficiencies, the signal is corrected for polar-angle-dependent selection and reconstruction efficiencies, and residual charge misidentification is corrected using the $p$--$q$ method. All the aforementioned methods are introduced in the following sections.

\subsubsection*{Template Subtraction and Efficiency Correction}

Background contributions are estimated using truth-level Monte Carlo templates, scaled by the corresponding selection efficiencies $\epsilon = N_{\text{sel}}/N_{\text{gen}}$ for each background $b$. The corrected signal yield in each bin is
\begin{equation}
N_{\text{signal}} = N_{\text{obs}} - \sum_b \epsilon_b N_b,
\end{equation}
After which, the signal is corrected for angular-dependent reconstruction and selection efficiency:
\begin{equation}
N_{\text{corr}} = \frac{N_{\text{signal}}}{\epsilon_{signal}}.
\end{equation}

\subsubsection*{Migration Correction (p-q method)}
Let $P_{\mathrm{chg}}$ be the probability that the sign we infer from the LPFO correctly reproduces the charge of the initial quark produced in the hard scattering. The probability of misidentification is then $Q_{\mathrm{chg}} = 1 - P_{\mathrm{chg}}$. Assuming that the jet-charge measurements in the two hemispheres are independent and symmetric, the number of events with compatible charges (opposite sign) in a given $|\cos\theta|$ bin can be written as
\begin{equation}
N_{\mathrm{acc}}(|\cos\theta|) 
= \left(P_{\mathrm{chg}}^{2} + Q_{\mathrm{chg}}^{2}\right)
\, N(|\cos\theta|),
\label{eq:pq_prob}
\end{equation}
where $N(|\cos\theta|)$ denotes the total number of events in the bin, and $N_{\mathrm{acc}}(|\cos\theta|)$ the number of events with opposite-sign reconstructed charges passing the selection.CPID

Once $P_{\mathrm{chg}}$ is determined from Eq.~\ref{eq:pq_prob}, the observed events in the forward and backward regions can be unfolded to recover the true distributions. The corrected event yields are obtained by inverting the corresponding $2\times2$ migration system:
\begin{equation}
\begin{aligned}
N_{\mathrm{corr}}(+|\cos\theta|) &= 
\frac{P_{\mathrm{chg}}^{2}\, N_{\mathrm{acc}}(\cos\theta>0)
- Q_{\mathrm{chg}}^{2}\, N_{\mathrm{acc}}(\cos\theta<0)}
{P_{\mathrm{chg}}^{4} - Q_{\mathrm{chg}}^{4}}, \\[6pt]
N_{\mathrm{corr}}(-|\cos\theta|) &= 
\frac{P_{\mathrm{chg}}^{2}\, N_{\mathrm{acc}}(\cos\theta<0)
- Q_{\mathrm{chg}}^{2}\, N_{\mathrm{acc}}(\cos\theta>0)}
{P_{\mathrm{chg}}^{4} - Q_{\mathrm{chg}}^{4}} .
\end{aligned}
\label{eq:pq_corr}
\end{equation}

This expression corresponds to the analytical inversion of the $2\times2$ charge-migration matrix under the assumption of identical charge-measurement performance in both jets.

\subsection*{Fit}
After applying the correction procedure, the resulting differential cross-section represents the parton-level distribution. This distribution is fitted using
\begin{equation}
\frac{d\sigma}{d\cos\theta} = S(1+\cos^2\theta) + A\cos\theta,
\end{equation}
neglecting the term proportional to $\sin^2\theta$, as it is strongly suppressed by the large Lorentz boost of the $s$-quarks. 
The fit is restricted to the barrel region, $|\cos\theta| < 0.8$, where the PFO information is not significantly affected by detector acceptance. 
The \AFB\ value is obtained by extrapolating this function to the full angular range. 
This method is motivated by the reduced statistics and degraded reconstruction performance in the very forward/backward regions ($|\cos\theta| > 0.8$).

\begin{figure}[!ht]
\centering
    \begin{tabular}{cc}
      \includegraphics[width=0.45\textwidth]{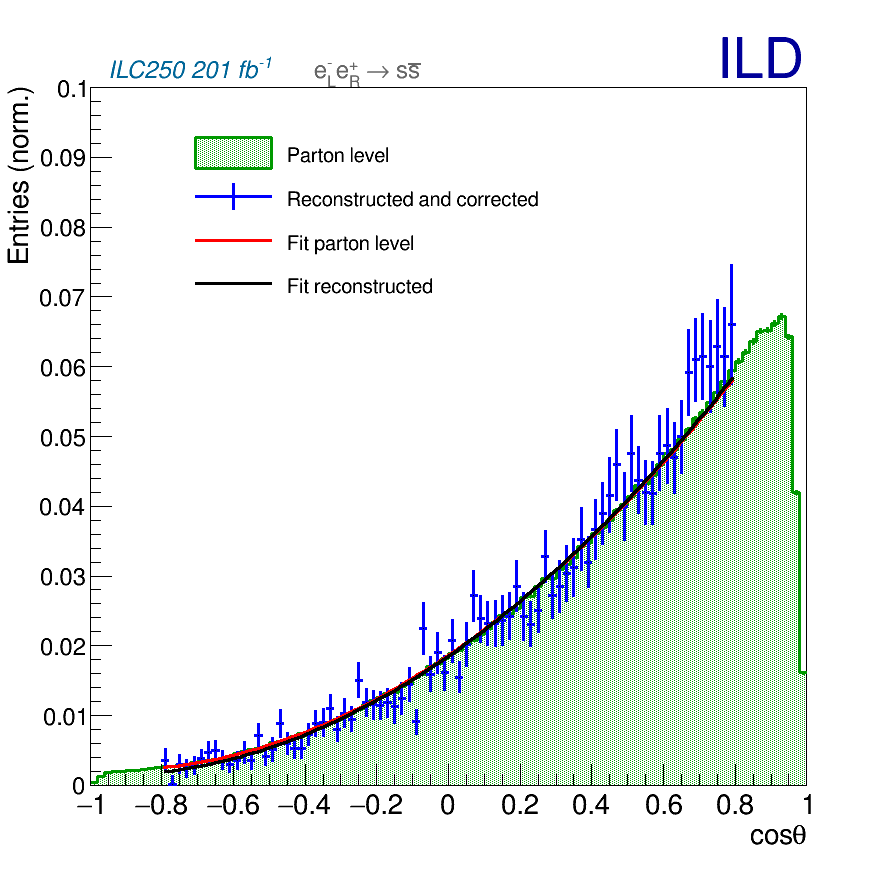} &
      \includegraphics[width=0.45\textwidth]{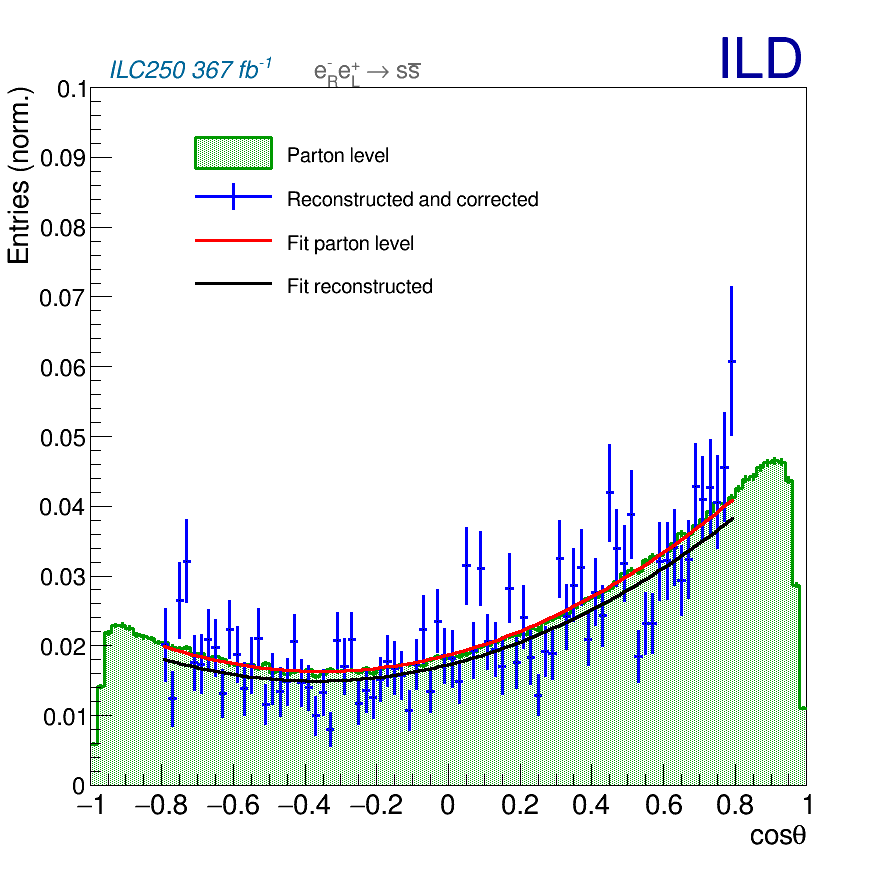}
    \end{tabular}
\caption{Comparison of the angular fits to the reconstructed and corrected data with respect to the parton level distributions. Plots for the \eLpR (left) and \eRpL (right) cases using \dEdx. \label{fig:fit}}
\end{figure}

An example of these fits for the baseline \dEdx case for \eLpR and \eRpL is shown in Fig.~\ref{fig:fit}, comparing parton-level and reconstructed distributions for both polarizations. A bias is observed in the \eRpL case, attributed to limited statistics in the samples, leading to an imperfect PID correction outside the fiducial region ($|\cos\theta| > 0.8$) and consequently distorting the event yield within the fitted range. A conservative estimate of the associated systematic uncertainties is included to account for this effect, as discussed in Sec.~\ref{sec:syst}.

\section{Improvements in PID}\label{sec:newPID}
\subsection*{Software improvements}\label{sec:softimprov}

The same analysis as described in Sec.~\ref{sec:previousanalysis} was repeated using the CPID framework (Ref.~\cite{Einhaus:2023oqy}). In this approach, the particle identification for each PFO is based on the likelihood outputs of BDT classifiers. In particular, cut~7 was modified such that a PFO is considered a kaon if $L_K > 0.5 \quad \text{and} \quad L_\pi < 0.7$, where $L_{K/\pi}$ are the kaon/pion likelihood outputs of CPID.
All other cuts are applied as in the \dEdx analysis.

\subsection*{Hardware improvements}
\label{sec:hardimprov}

Two scenarios are considered to evaluate the potential gain from improved PID. A pixel-based TPC is expected to improve kaon/pion separation by $\sim30$--$40\%$ for track momenta between $3$ and $50$~GeV~\cite{LCTPC:2022pvp} by using \textit{Cluster Counting} (\dNdx). Since \dNdx reconstruction is not yet available in ILD, its effect is emulated by narrowing the PID likelihood distributions, following Refs.~\cite{Marquez:2023guo,Irles:2024ipg}. In the \dNdx scenario, the standard deviation of the $k$-distance is reduced by $25\%$ (corresponding to $\sim33\%$ improvement in separation), while in the \textit{PerfectTPC} case it is reduced by $99\%$, representing near-ideal PID. The baseline \dEdx analysis flow is applied in both cases for direct comparison.

\section{Estimation of systematic uncertainties}\label{sec:syst}
Systematic uncertainties are evaluated by propagating the uncertainties on the selection efficiencies for both signal and background through the template subtraction procedure:

\begin{enumerate}
\item For each toy, the template subtraction is repeated, yielding a set of corrected distributions.

\item For each toy, the full fit is performed, resulting in a distribution of $A_{\mathrm{FB}}$ values.

\item The impact of the efficiency uncertainties on $A_{\mathrm{FB}}$ is evaluated from the distribution of toy results. The spread of the distribution, quantified by its root-mean-square (RMS), reflects the sensitivity to the fluctuations, while the mean shift with respect to the reference value accounts for a potential bias induced by the procedure. A conservative estimate of the systematic uncertainty is then defined as
\begin{equation}
    \Delta_{A_{\mathrm{FB}}} = \sqrt{\Delta_{\text{disp}}^2 + \Delta_{\text{rms}}^2},
\end{equation}
where $\Delta_{\text{disp}}$ is the mean shift and $\Delta_{\text{rms}}$ is the RMS of the distribution.
\end{enumerate}

There are other sources of systematic uncertainties that haven't been considered here, such as polarization in the initial state or hemisphere correlation in the detector, but, as can be seen in Ref.~\cite {Irles:2023ojs}, their size is minimal compared to the template subtraction.

\section{Results for ILC250 and LCF@CERN250}\label{sec:results}

\begin{figure}[!ht]
\centering
    \begin{tabular}{cc}
      \includegraphics[width=0.45\textwidth]{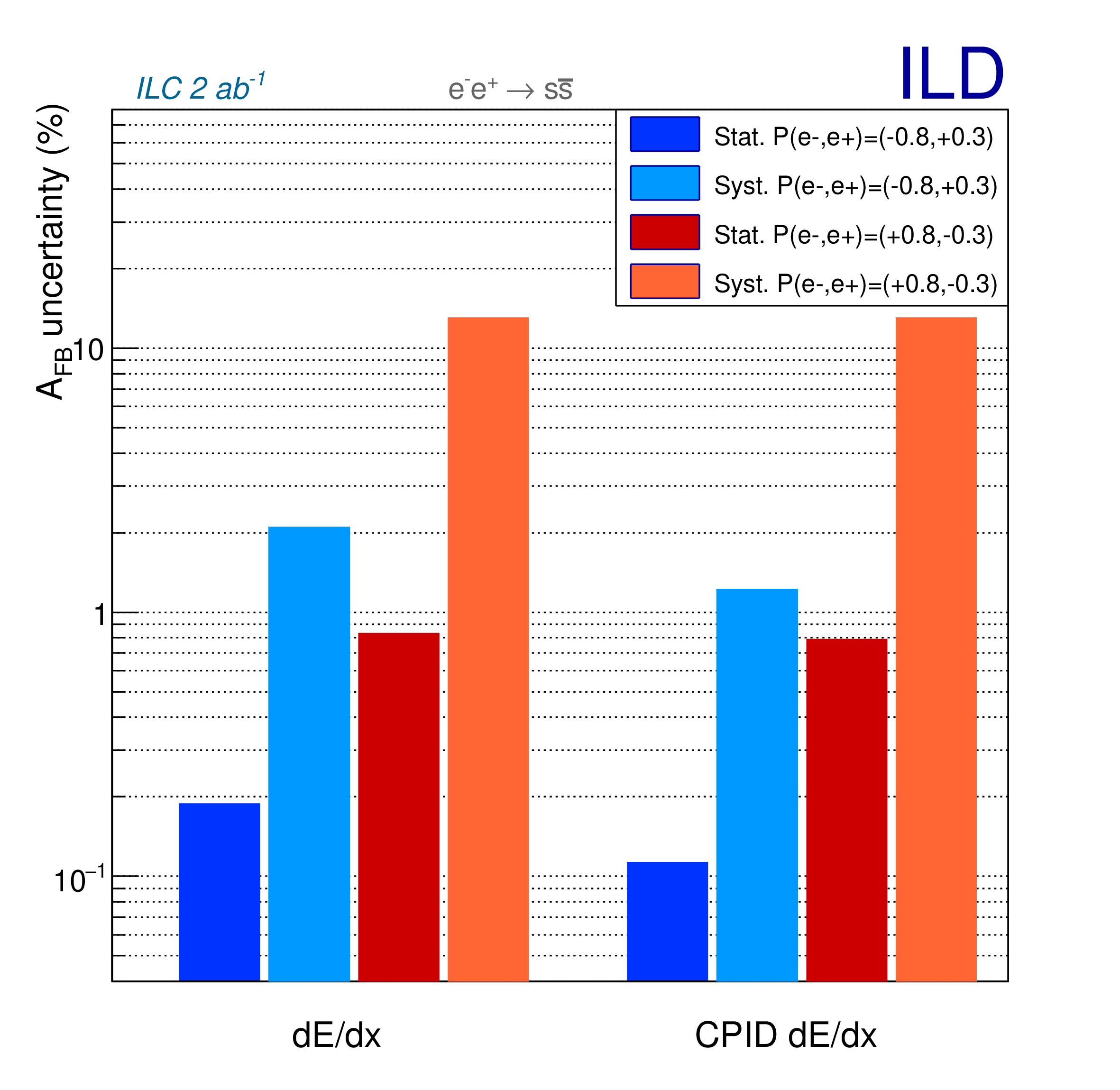} &
      \includegraphics[width=0.45\textwidth]{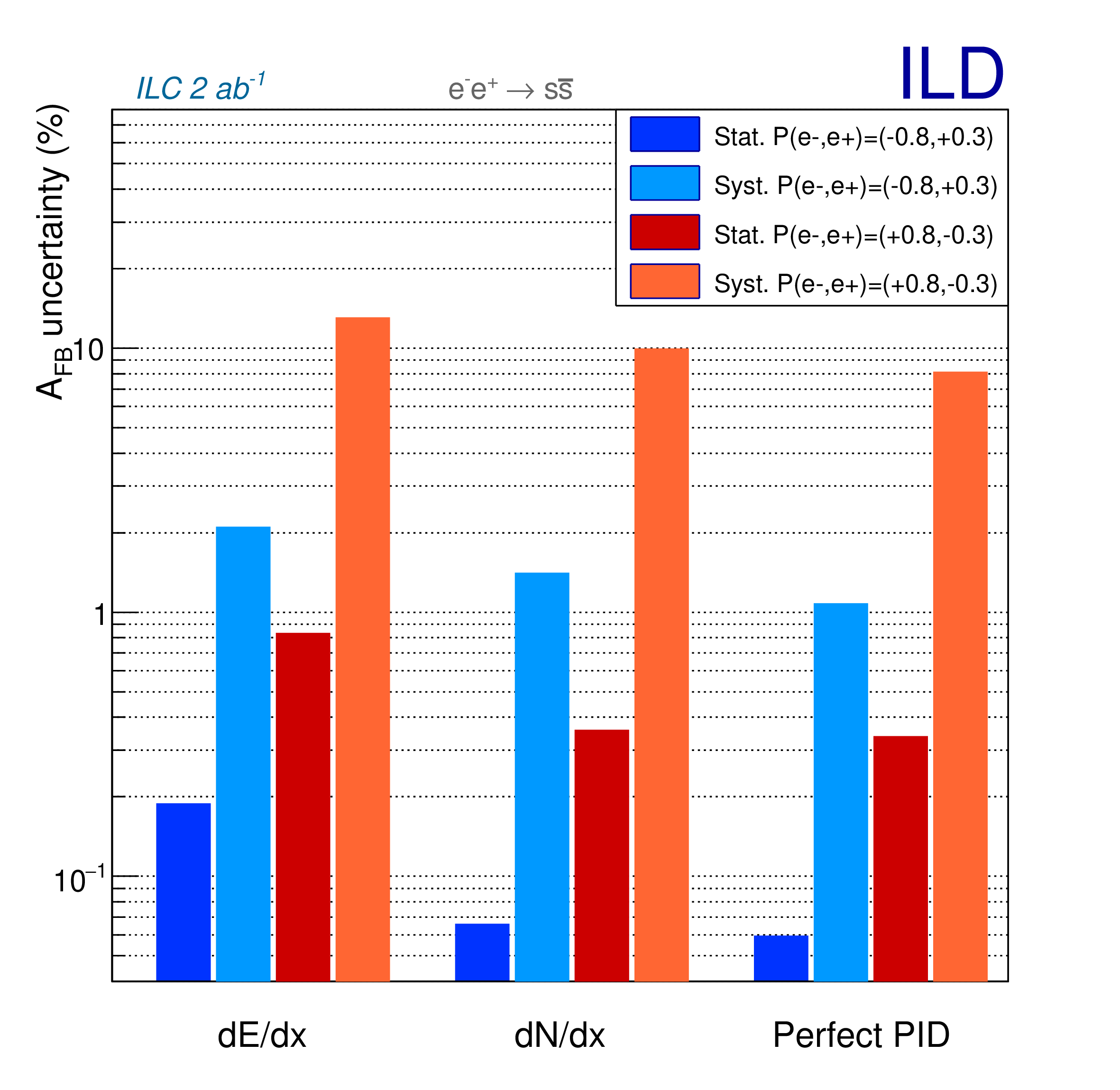}
    \end{tabular}
\caption{Expected uncertainties for \ssbar in the ILC250. Left: Comparing different software approaches (\dEdx vs CPID \dEdx). Right: Comparing different hardware settings (\dEdx vs \dNdx in a pixel TPC vs An ideal TPC) \label{fig:resultsILC}}
\end{figure}

\begin{figure}[!ht]
\centering
    \begin{tabular}{cc}
      \includegraphics[width=0.45\textwidth]{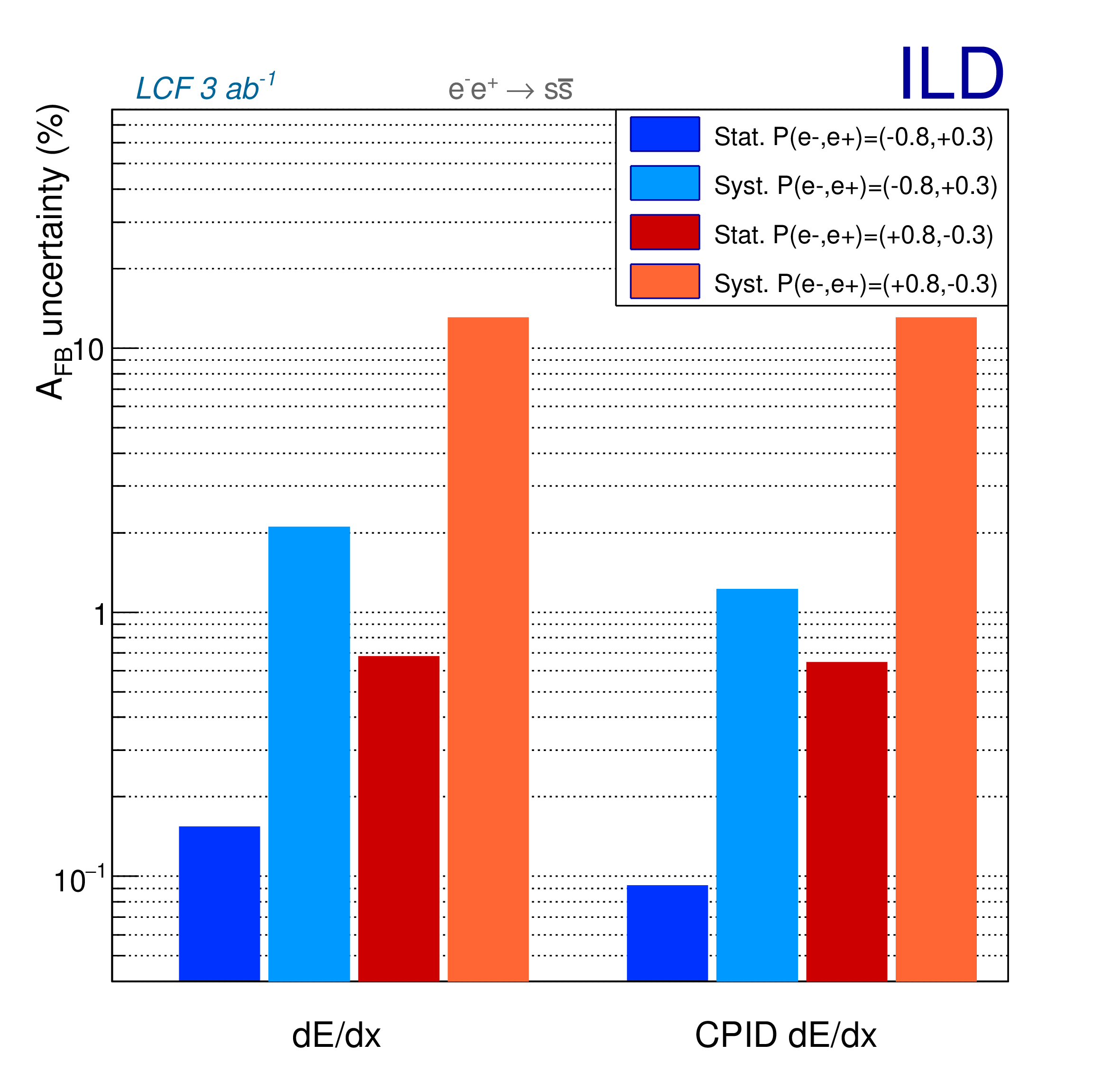} &
      \includegraphics[width=0.45\textwidth]{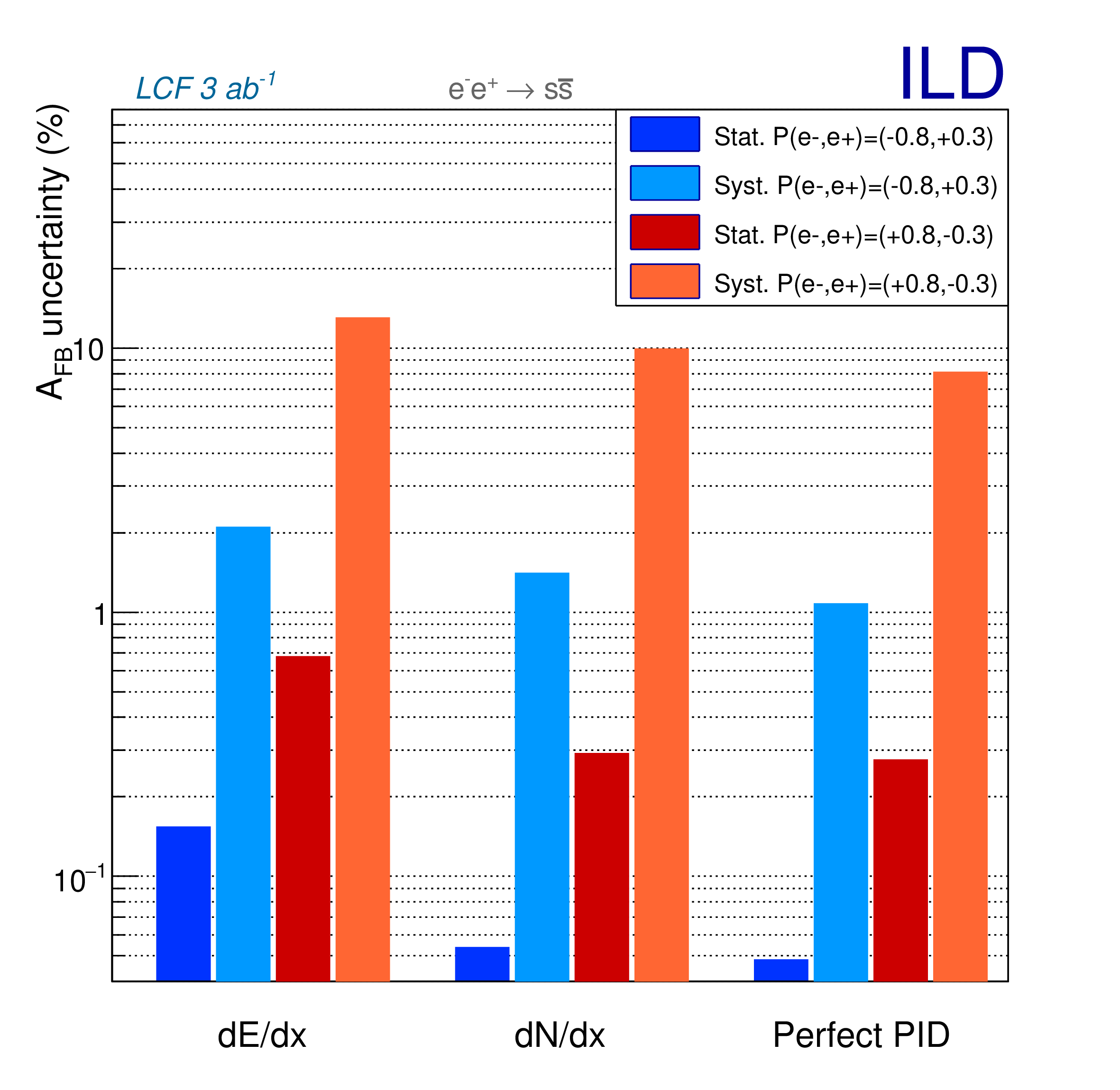}
    \end{tabular}
\caption{Expected uncertainties for \ssbar in the LCF@CERN250. Left: Comparing different software approaches (\dEdx vs CPID \dEdx). Right: Comparing different hardware settings (\dEdx vs \dNdx in a pixel TPC vs An ideal TPC) \label{fig:resultsLCF}}
\end{figure}

Figures~\ref{fig:resultsILC} and~\ref{fig:resultsLCF} summarise \AFB results for ILC250 and LCF@CERN250. Statistical uncertainties are scaled to expected integrated luminosities. Improvements from CPID and hardware upgrades reduce uncertainties, with \textit{PerfectTPC} providing a lower benchmark. As can be seen in the plots, systematic uncertainties dominate for all cases, and both software (CPID) and hardware (dN/dx pixel TPC) offer potential to improve the measurements. The best improvements are observed in the \eLpR channels, where CPID yields a $\sim 40\%$ reduction of both statistical and systematic uncertainties, and \dNdx yields a $\sim 66\%$ reduction of statistical and $\sim 30\%$ of systematic uncertainties.

%\section{Possible impact on GHU phenomenology}\label{sec:ghu}
%
%Figure~\ref{fig:resultsGHU} shows the potential impact of improved $s\bar{s}$ %forward-backward asymmetry measurements on discriminating \textit{Gauge-Higgs %Unification} (GHU) scenarios. The plots follow the methodology of %Ref.~\cite{Irles:2024ipg}, using projected experimental uncertainties from %Ref.~\cite{Altmann:2025feg} as input. These uncertainties are indicative estimates %for the \bbbar\ and \ccbar\ channels using the \textit{Particle Transformer} (ParT); %a dedicated analysis will be required. The figures demonstrate that even modest %improvements in $s\bar{s}$ \AFB can enhance sensitivity to GHU parameters.
%
%\begin{figure}[!ht]
%\centering    
%\begin{tabular}{cc}
%      \includegraphics[width=0.45\textwidth]{figures/GHU_1percent_PRE.png} &
%      \includegraphics[width=0.45\textwidth]{figures/GHU_1permil_PRE.png}
%    \end{tabular}
%\caption{Expected impact in the discrimination power for GHU models assuming a 1\% %(left) or 0.1\% (right) uncertainty for \ssbar. \label{fig:resultsGHU}}
%\end{figure}

\section{Conclusions}

This study demonstrates the improvements achievable in \ssbar forward–backward asymmetry measurements through enhanced particle identification. Upgrading from standard \dEdx to emulated \dNdx, along with applying CPID likelihood-based selections, results in a clear reduction in both statistical and systematic uncertainties.

The present analysis, based on a cut-based approach, provides a robust baseline for assessing these gains, although it does not fully exploit the available kinematic and PID information. Ongoing studies using the \textit{Particle Transformer} (ParT) framework (Refs.~\cite{qu2024particletransformerjettagging,tagami2024applicationparticletransformerquark}) aim to address these limitations by enabling more powerful multivariate selections, with the potential for further improvements in sensitivity.

\section*{Acknowledgements}
We thank the LCC generator and ILD software working groups for supplying the simulation and reconstruction tools, as well as for producing the Monte Carlo samples employed in this study. We also thank A. Irles and Y. Okugawa for providing the software from the previous analyses, as well as very useful inputs for new developments.
This work was funded by the French-German ANR-DFG project CALO5D under the Grant Agreement ANR-23-CE31-0023.

%\printbibliography[title=References]
\bibliography{sn-bibliography}

\newpage
\section*{Appendix}\label{sec:Appendix}

\subsection*{Efficiencies for the \dEdx case}

\begin{table}[!ht]
\centering
\begin{tabular}{c|ccccc}
\multicolumn{1}{c}{} & \multicolumn{5}{c}{Efficiency (\%)}\\ 
\multicolumn{1}{c}{} & dd & uu & ss & cc & bb\\ \hline
+ Cut 1  & 93.9   & 93.9   & 93.1   & 69.3   & 2.11  \\ 
+ Cut 2  & 91.8   & 91.6   & 90.9   & 14.0   & 1.35  \\ 
+ Cut 3  & 91.8   & 91.6   & 90.9   & 14.0   & 1.35  \\ 
+ Cut 4  & 44.9   & 51.8   & 42.3   & 4.01   & 0.07  \\ 
+ Cut 5  & 38.2   & 43.9   & 35.9   & 3.35   & 0.06  \\ 
+ Cut 6  & 36.8   & 42.4   & 34.0   & 3.10   & 0.05  \\ 
+ Cut 7  & 2.38   & 2.91   & 4.80   & 0.22  & $<$0.01  \\ 
+ Cut 8  & 0.29  & 0.46  & 0.63  & 0.04 & $<$0.01  \\ 
+ Cut 9  & 0.17  & 0.33  & 0.48  & 0.02 & $<$0.01  \\ 
\end{tabular}
\caption{Efficiencies per cut in the \eLpR case.}
\label{tab:eff_eLpR}
\end{table}

\begin{table}[!ht]
\centering
\begin{tabular}{c|ccccc}
         & dd & uu & ss & cc & bb\\ \hline
+ Cut 1  & 4530066 & 6758044 & 4491002 & 4985282 & 101577\\ 
+ Cut 2  & 4425360 & 6596732 & 4382552 & 1010825 & 64977\\ 
+ Cut 3  & 4425360 & 6596732 & 4382552 & 1010825 & 64977\\ 
+ Cut 4  & 2165749 & 3725928 & 2040522 & 288641 & 3549\\ 
+ Cut 5  & 1841640 & 3162916 & 1730440 & 241304 & 2734\\ 
+ Cut 6  & 1776736 & 3051260 & 1641495 & 222918 & 2316\\ 
+ Cut 7  & 114986 & 209620 & 231436 & 15960 & 75\\ 
+ Cut 8  & 14102 & 33066 & 30317 & 3150 & 38\\ 
+ Cut 9  & 8066 & 23643 & 22928 & 1601 & 23\\ 
\end{tabular}
\caption{Number of events per cut in the \eLpR case.}
\label{tab:N_eLpR}
\end{table}

\begin{table}[!ht]
\centering
\begin{tabular}{c|ccccccccc}
Purity & Cut 1 & Cut 2 & Cut 3 & Cut 4 & Cut 5 & Cut 6 & Cut 7 & Cut 8 & Cut 9 \\ \hline
        & 21.5\% & 26.6\% & 26.6\% & 24.8\% & 24.8\% & 24.5\% & 40.5\% & 37.6\% & 40.8\% \\
\end{tabular}
\caption{Purity per cut in the \eLpR case.}
\label{tab:N_eLpR}
\end{table}

\begin{figure}[!ht]
\centering
      \includegraphics[width=1.\textwidth]{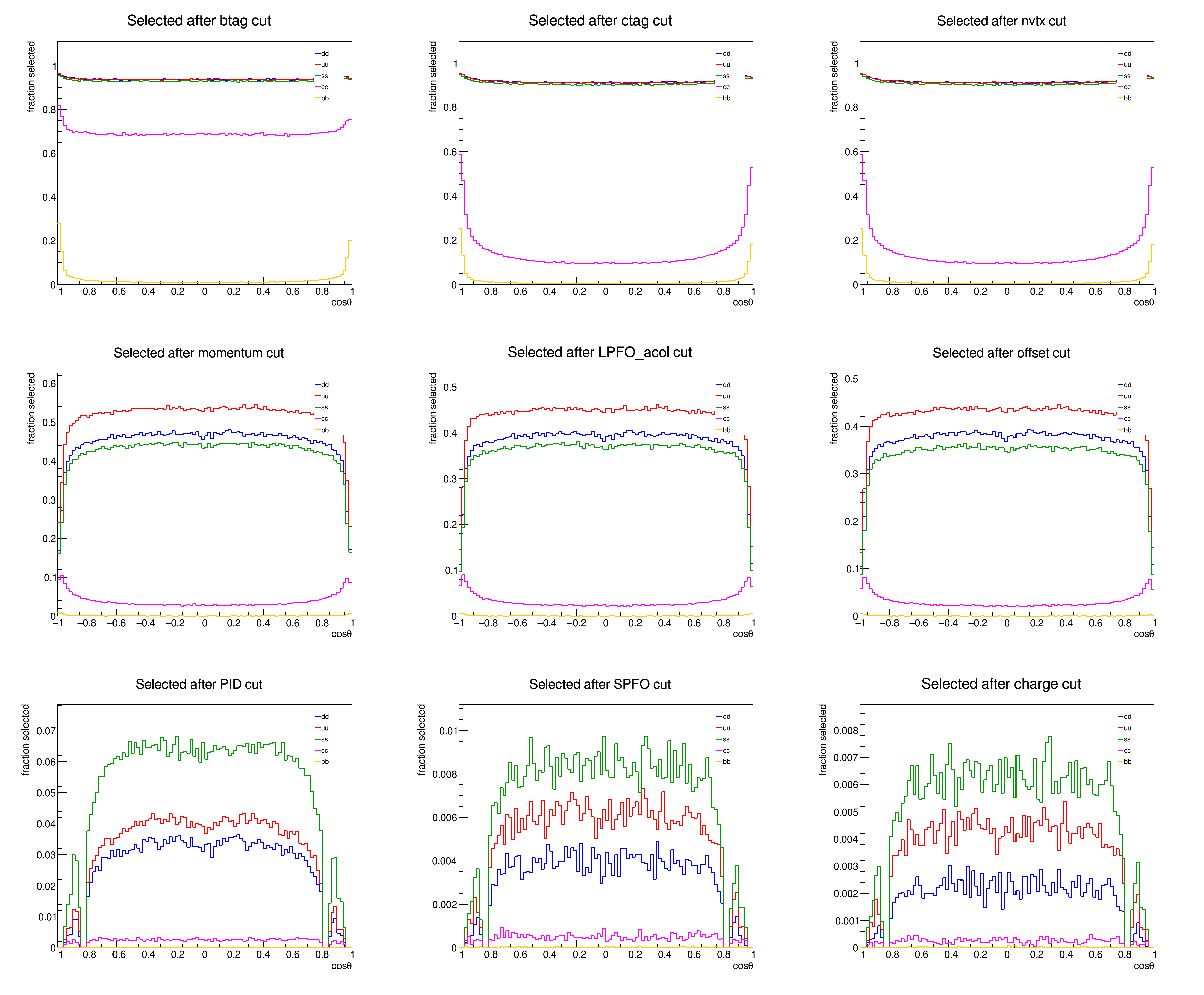}
\caption{Efficiencies per cut in the \eLpR case. These plots are cumulative. \label{fig:eff_eLpR}}
\end{figure}

\begin{table}[!ht]
\centering
\begin{tabular}{c|ccccc}
\multicolumn{1}{c}{} & \multicolumn{5}{c}{Efficiency (\%)}\\ 
\multicolumn{1}{c}{} & dd & uu & ss & cc & bb\\ \hline
+ Cut 1  & 94     & 93.9   & 93.2   & 69.3   & 3.45  \\ 
+ Cut 2  & 92     & 91.7   & 91.1   & 14.6   & 2.46  \\ 
+ Cut 3  & 92     & 91.7   & 91.1   & 14.6   & 2.46  \\ 
+ Cut 4  & 38.2   & 50.4   & 35.9   & 3.92   & 0.07  \\ 
+ Cut 5  & 32.4   & 42.7   & 30.4   & 3.30   & 0.06  \\ 
+ Cut 6  & 31.3   & 41.2   & 28.8   & 3.04   & 0.05  \\ 
+ Cut 7  & 1.99   & 2.84   & 4.07   & 0.22   & $<$0.01  \\ 
+ Cut 8  & 0.26   & 0.44   & 0.53   & 0.04   & $<$0.01  \\ 
+ Cut 9  & 0.15   & 0.32   & 0.40   & 0.02   & $<$0.01  \\ 
\end{tabular}
\caption{Efficiencies per cut in the \eRpL case.}
\label{tab:eff_eRpL}
\end{table}

\begin{table}[!ht]
\centering
\begin{tabular}{c|ccccc}
         & dd & uu & ss & cc & bb\\ \hline
+ Cut 1  & 1117388 & 3161191 & 1108019 & 2334710 & 41036\\ 
+ Cut 2  & 1093149 & 3086813 & 1083110 & 493021 & 29290\\ 
+ Cut 3  & 1093149 & 3086813 & 1083110 & 493021 & 29290\\ 
+ Cut 4  & 454493 & 1696292 & 426315 & 132022 & 825\\ 
+ Cut 5  & 385612 & 1438706 & 361736 & 110969 & 653\\ 
+ Cut 6  & 371649 & 1388972 & 342885 & 102527 & 551\\ 
+ Cut 7  & 23627 & 95543 & 48350 & 7354 & 15\\ 
+ Cut 8  & 3035 & 14681 & 6259 & 1491 & 7\\ 
+ Cut 9  & 1727 & 10667 & 4743 & 781 & 2\\ 
\end{tabular}
\caption{Number of events per cut in the \eRpL case.}
\label{tab:N_eRpL}
\end{table}

\begin{table}[!ht]
\centering
\begin{tabular}{c|ccccccccc}
Purity & Cut 1 & Cut 2 & Cut 3 & Cut 4 & Cut 5 & Cut 6 & Cut 7 & Cut 8 & Cut 9 \\ \hline
        & 14.3\% & 18.7\% & 18.7\% & 15.7\% & 15.7\% & 15.5\% & 27.6\% & 24.6\% & 26.5\% \\
\end{tabular}
\caption{Purity per cut in the \eRpL case.}
\label{tab:N_eRpL}
\end{table}

\begin{figure}[!ht]
\centering
      \includegraphics[width=1.\textwidth]{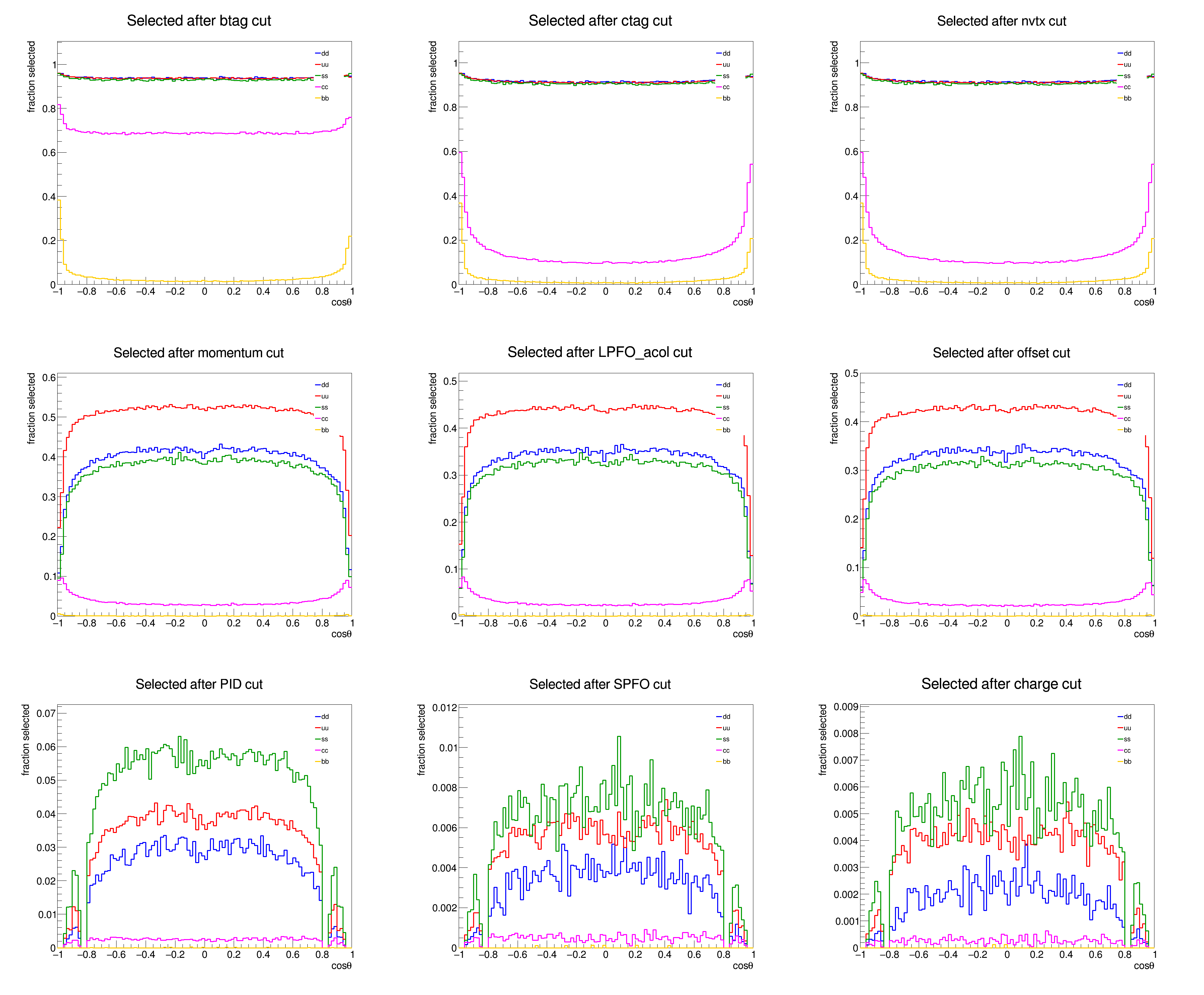}
\caption{Efficiencies per cut in the \eRpL case. These plots are cumulative.  \label{fig:eff_eRpL}}
\end{figure}

\end{document}